# An empirical study of spatial and transpatial social networks using Bluetooth and Facebook

Vassilis Kostakos

University of Madeira
and
Carnegie Mellon University

vassilis@cmu.edu

**Abstract**

This study provides insights into the quantitative similarities, differences and relationships between users' spatial, face-to-face, urban social networks and their transpatial, online counterparts. We explore and map the social ties within a cohort of 2602 users, and how those ties are mediated via physical co-presence and online tools. Our analysis focused on isolating two distinct segments of the social network: one mediated by physical co-presence, and the other mediated by Facebook. Our results suggest that as a whole the networks exhibit homogeneous characteristics, but individuals' involvement in those networks varies considerably. Furthermore this study provides a methodological approach for jointly analysing spatial & transpatial networks utilising pervasive and ubiquitous technology.

Keywords: Social network, Facebook, Bluetooth, spatial and transpatial.

## Introduction

In recent years network science has emerged as an important approach to analysing and modelling dynamic and static systems, with prime examples of such systems being social networks. Coupled with the growing data collecting capabilities of pervasive and ubiquitous systems, network analysis can elicit important insights into the structure and behaviour of users and their relationships, both in online and urban environments. This study offers novel insights into the quantitative similarities, differences and relationships between users' spatial, face-to-face, urban social networks and their transpatial, online counterparts. We conducted this quantitative study to explore and map the social ties between a cohort of 2602 users, and how those ties are mediated via physical co-presence and online tools. To collect information about our cohort's spatial and transpatial social networks we relied on proxies to those networks. As a proxy to spatial networks we relied on the use of Bluetooth to record physical co-presence between those users. On the other hand, we used explicit friendship ties on Facebook as a proxy to transpatial networks.

### Spatial and transpatial networks

Social networks are a significant resource that humans have at their disposal. Members of our social networks support each other in many important aspects of their lives including in family and career matters. Historically, social networks have been strongly spatial with much of pre-history comprising of small groups of hunter-gatherers living and travelling together. The emergence of villages and eventually cities gave rise to neighbourhoods where people living in physical proximity formed tight-knit communities. More recently, strongly situated daily activities have given rise to social networks: the workplace, the school and the cafe are all examples of locations giving rise to and fostering social networks.

In parallel, advances in communication technology, such as pen-pal writing, the telephone, email, forums, instant messaging, video conferencing and social networking websites helped brake away from the limitation of location-bound social networking and allowed for meaningful social relationships to become transpatial. Transpatial relationships complement and can potentially replace the more traditional spatial, face-to-face social ties. A substantial amount of research has focused on understanding how these technological advances have changed society, the way we communicate and the way we support each other [e.g. Castells, 1995]. While dystopian views of the future described physically isolated individuals communicating solely via computers and other technological means, today urbanisation levels have surpassed 50% for the first time in history [United Nations, 2007], hence providing unprecedented opportunities for spatial networking. Yet at the same time communication technologies have made important advances and achieved widespread penetration, with mobile phones and the internet leading the way in the generation and maintenance of transpatial social networks.

Pervasive and ubiquitous systems' primary focus on the human makes them a prime example of technology that can help users manage their access to social networks, both spatial and transpatial. Pervasive technology can improve and augment face-to-face social interactions by providing contextual information and memory aids, and at the same time offer the means to improve communication between distant users thus helping them maintain transpatial ties. Yet, to design such technology it is important to understand the relationship between these two types of social networks. For example, how do spatial and transpatial ties contribute to the overall structure of a user's social network? To what extend does technology have the potential to completely overtake users' face-to-face social interactions? More crucially, are user's roles and involvement in spatial and transpatial networks related in any way? The subtle differences, similarities and dynamics between spatial and transpatial social networks can potentially determine the success or failure of systems that aim to provide a grand unified means of access to social networks.

The focus of this study is to uncover the similarities and differences between users' spatial and transpatial networks, and more importantly to understand how these two types of social networks combine to create a single social circle for each user. A considerable portion of our work focuses on the structural aspects of users' social network and their respective analysis.

In the next section we provide a brief overview of the technology we used to collect our data. We then summarise previous work that is related to the present study. Subsequently we describe in detail the type of data we actually collected from our cohort, and follow up with a presentation of the results of our study. Finally we provide a discussion of our results.

## The Cityware Application

The study reported here was conducted using the Cityware application (see [Kostakos & O'Neill, 2008] for a complete description of the system). Cityware is a massively distributed application, spanning both the online and physical worlds. Its architecture uniquely allows it to expand and contract in real time, while also enabling live data analysis. The main components of the platform are: people's Bluetooth-enabled devices, Cityware nodes, Cityware servers, Facebook servers, and a Facebook application. An overview of this architecture is shown in Figure 1.

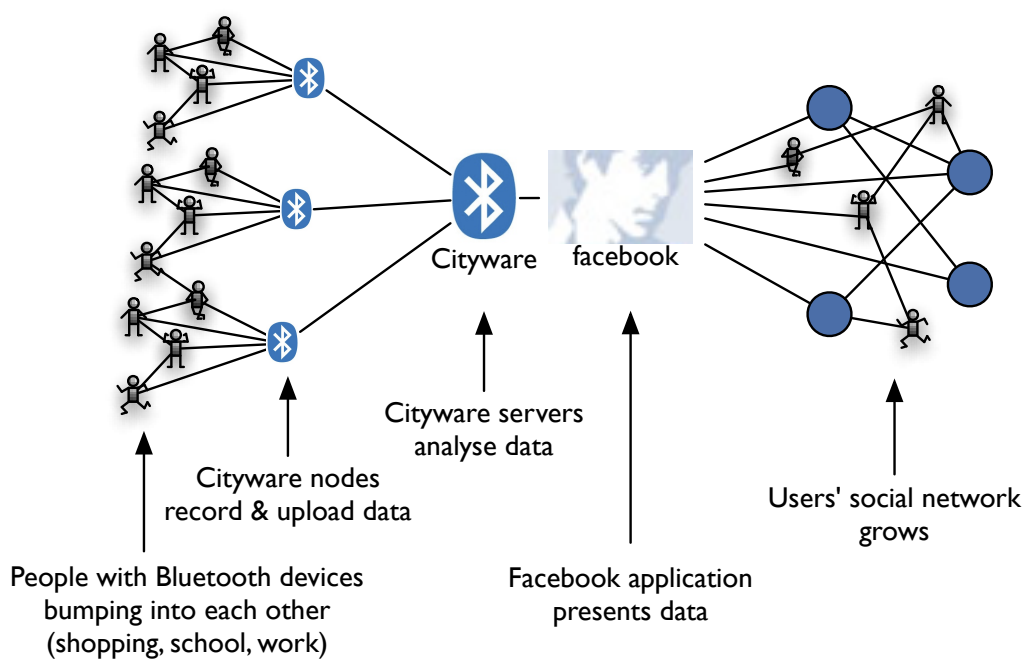

**Figure 1. Overview of the Cityware platform.**

User-run Cityware nodes are Bluetooth-enabled PCs that collect and upload information about nearby people's presence and encounter in physical space. Since Bluetooth enables the unique identification of individual devices, we can measure the behaviour of individual devices (hence users), as well as the relationship between pairs of devices (hence pairs of users). Using Facebook as a front-end to our application, users can register their personal Bluetooth devices. Hence, for each registered user our system knows their unique Bluetooth ID and their unique Facebook profile ID. We note that a user can register multiple Bluetooth devices, hence a Facebook ID can be associated with multiple Bluetooth IDs. In addition, we should point out that we cannot know whether users are being truthful about their claim to ownership of a Bluetooth device. Our assumption is that users indeed own and use the Bluetooth devices they claim to. Finally, we point out that our system is opt-in, hence users willingly associate Bluetooth devices with their Facebook profile.

Before describing the type of data we collected using this system, we first give an overview of previous work that is strongly related to our study.

## Related work

The formalised study of network graphs arguably begun by Euler's famous solution to the Seven Bridges of Königsberg problem in 1736 [Biggs et al., 1986]. In his solution, Euler represented the four landmasses and seven bridges of Königsberg, now Kaliningrad, as four nodes and seven links respectively. Thus, he was able to prove that no route crosses each bridge only once. Graph theory has greatly advanced every since, mostly focusing on mathematical proofs and theorems on graph topology, trees and cycles.

While graphs have been used to explore relationships between social entities for over a century, it was not until the 1950's that this became a systematic, and ultimately scientific process. Some of the first studies to engage in social network analysis are the kinship studies of Elizabeth Bott [Bott, 1957] and the urbanisation studies pioneered by Max Gluckman in Zambia [Gluckman & Aronoff, 1976]. Similarly, Granovetter's work [1973] lay the foundations for the small world hypothesis, suggesting that everyone is within six degrees of separation, while Wellman's work gave some evidence of how large-scale social changes have affected the nature of personal communities and the support they provide [1979]. Since then, social network analysis has moved from being a suggestive metaphor to becoming an analytic approach, with its own theories and research methods. In the 1970's, Freeman developed a multitude of metrics for analysing social and communication networks [e.g. 2004], thus boosting commercial interest in the area due to companies aiming to optimise their procedures and operations. In the last decade, the identification of mathematical principles such as the small-world and scaling phenomena [Barabasi & Albert, 1999; Watts & Strogatz, 1998], underpinning many natural and man-made systems, have sparked further interest in the study of networks.

The systems design community has also been interested in the study of social networks as well as online social networks. Typical research topics in the area include the effect of social engagement on behaviour [e.g. Millen & Patterson, 2002], the issue of identity and projected identity [Lee & Nass, 2003], as well as the design of socio-technical systems [Herrmann et al., 2004]. The recent proliferation of online social networking system such as Facebook, Dodgeball and MySpace, has provided researchers with platforms for carrying out research into online social behaviour, and a journal devoted to this topic (http://www.elsevier.com/locate/socnet). In the Urban Computing domain, such studies have looked at the effect of social incentives and contextual information on the use of public transportation [Booher et al., 2007], the relationship between users' online profiles and their online behaviour [Lampe et al., 2007], the various trust issues that emerge from using such systems [Riegelsberber & Vasalou, 2007], how such systems can help strengthen neighbourhoods [Foth, 2006], and the development of systematic grounds to base our designs [Kostakos et al., 2006].

As is evident from the above research, to make inferences from online behaviour datasets researchers still have to collect data from the real world and relate it to the online data. Thus, while social networking websites make it easy to capture large amounts of data, researchers still need to employ interviews, focus groups, questionnaires, or any other method that enables them to relate online with real world data.

The study reported here is itself is an example of the importance of Pervasive and Ubiquitous technology as a research tool, and specifically of the Cityware platform which aims to bridge the gap between online and physical social networks. It allows users and researchers to explore an amalgamation of online and physical social networks. The key strength of the Cityware platform is that it allows the collection of vast amounts of quantitative data, both from the online and real worlds, which is immediately linked, synchronised, and available for further analysis. Furthermore, this platform enables both end users and researchers to gain a better understanding of the relationship between spatial and transpatial social networks. For a detailed description of the architecture of our platform, the types of data it makes available to users and researchers, and the typical user-oriented scenarios that in enables, the reader is referred to [Kostakos & O'Neill, 2008].

A number of projects have carried out research relating to Cityware's physical co-presence and tracking across cities. The Reality Mining project collected proximity, location and activity information, with proximity nodes being discovered through periodic Bluetooth scans and location information by cell tower IDs. Several other groups have performed similar studies [Eagle & Pentland, 2006; Balazinska & Castro, 2003; Chaintreau et al., 2006; McNett & Voelker, 2005; Nicolai et al., 2005]. Most of these, such as [Balazinska &

Castro, 2003] and [Nicolai et al., 2005], use Bluetooth to measure mobility, while others, such as [Chaintreau et al., 2006] and [McNett & Voelker, 2005;], rely on WiFi. The duration of experiments varies from 2 days to over 100 days, and the numbers of participants vary from 8 to over 5000 (see the Haggle project). The Crawdad database provides extensive traces, which are useful for the validation of forwarding algorithms and routing protocols that operate through learning characteristics of node mobility. In or work, our datasets consist of more than 150000 participants over two years of data at the time of writing.

Furthermore, a number of projects measure various aspects of technology and users on a large scale. For example, the MetroSense project explores the use of people-centric sensing with personal as well as consumer oriented sensing applications such as Nike+, and sensor-enabled mobile phone applications. Sensing can potentially cover a campus, metropolitan area, or a whole city, with many potential applications such as noise mapping and pollution mapping. The Pervasive Mobile Environmental Sensor Grids (MESSAGE) project aims to collect data at a metropolitan scale through smart phones carried by cyclists, cars, and pedestrians monitoring carbon dioxide values to control traffic in the city of Cambridge. Similarly, the urban sensing project CENS seeks to develop cultural and technological approaches for using embedded and mobile sensing to invigorate public space and enhance civic life.

**Data collection**

For the study reported here we collected and analysed data from a cohort of 2602 users. Using Bluetooth technology, we collected data on the physical copresence of the cohort over a period of one month in March 2007. This data was collected by users of the Cityware application. Hence, the hardware which collected this data in most cases was not ours, but rather the software was. Since we only had control over the software used to collect data, all our users' hardware most probably was not identical, and therefore we should bear in mind that the data was collected by Bluetooth transceivers of varying signal strength, as as well as variable positioning in relation to the flow of pedestrians being observed. The effects of varying signal strength does not significantly effect our data since stronger signal strength simply results in broader coverage of our system. It is not yet clear the effect that positioning of the hardware can have.

At the end of the Bluetooth data-collecting period we recorded the explicit friendship relationships between the same set of users on Facebook. Due to Facebook's privacy architecture, a query regarding the friendship status between two users may return as

- positive (i.e. they are indeed friends)
- negative (i.e. they are not friends)

- null (i.e. due to users' privacy settings we cannot know if they are friends or not).

Using this double-pronged approach, we collected two types of data for the same social network. Bluetooth enables us to calculate co-presence between individuals in our cohort by considering instances when two individuals where detected by a Cityware node at the same place and the same time. On the other hand, Facebook gives us insight into explicit friendship ties between the same set of users.

**Results**

### Data coding

The collected data was converted to social network graphs as follows. To analyse the Bluetooth dataset we represented each user as a node, and we connected those individuals who had been physically co-present at some point during our observation. In cases where users owned multiple Bluetooth devices, then an encounter with any of their devices would result in a link.

Next, we generated a graph from our Facebook dataset by representing each user as a node, and linking together users who were friends on Facebook. In cases where due to users' privacy settings a friendship tie returned as null, we treated this tie as a non-friendship tie.

Finally, we fused these two distinct social networks as follows: each member of our cohort was represented as a node, and we linked together nodes that were linked in either the Bluetooth or Facebook graphs. Hence, the fused network had three types of links: those that resulted from a Bluetooth encounter, those that resulted from a Facebook friendship, and those that resulted from both a Bluetooth encounter and a Facebook friendship.

A visual representation of these three social networks is shown in the figure below. The node and edge colour varies from blue to red, indicating low betweenness (blue) or high betweenness (red).

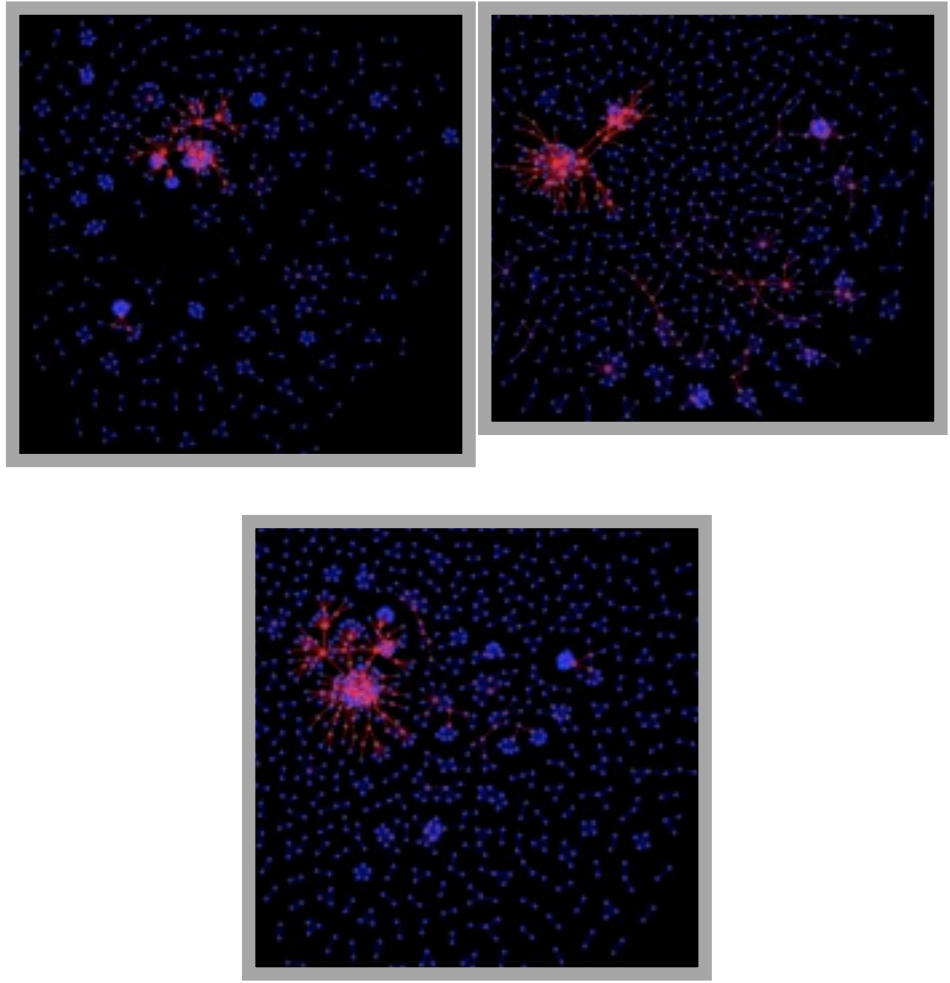

**Figure 1. Social networks derived from Bluetooth (top), facebook (middle), and the fusion of Bluetooth & Facebook (bottom) fused network. The node and edge colour varies from blue to red, indicating low betweenness (blue) or high betweenness (red).**

## Structural metrics

In Table 1 we present some structural metrics for the three graphs we derived. In addition to the size and number of edges of each graph, we present their density (portion of possible edges being instantiated), the size of the largest connected component of each graph (core), the average number of links that each node has (degree), the longest shortest-path of each graph (diameter), the average shortest-distance between all pairs of nodes ($\lambda$), and each graph's transitivity (clustering coefficient).

| Network | Size | Edges | Density | Core | $k$ | $\lambda_{max}$ | $\lambda$ | $C$ |
|---|---|---|---|---|---|---|---|---|
| Bluetooth | 2602 | 997 | 0.015% | 138 | 0.776 | 8 | 3.72 | 0.56 |
| Facebook | 2602 | 844 | 0.012% | 102 | 0.649 | 9 | 3.47 | 0.41 |
| Fused | 2602 | 1437 | 0.021% | 219 | 1.105 | 9 | 4.25 | 0.48 |

**Table 1. Structural properties of the obtained network and some of its subsets. For each subset we show size of the graph, number of edges in the graph, density of edges, size of largest component (core), average degree (k), diameter of largest component (λmax), average path length (λ), and average clustering coefficient (C).**

### Cluster sizes in the social networks

A further result we derived from our data was the size of the clusters that existed in each of the three graphs. All three graphs consisted of a number of distinct components, and cluster size refers to the size of these components. The largest component in each graph is referred to as the core. Figure 2 shows the probability distribution of cluster size for each of the three social networks we generated. The mean cluster size for the Bluetooth network was 1.20, for Facebook 1.27, and for the fused network 1.36.

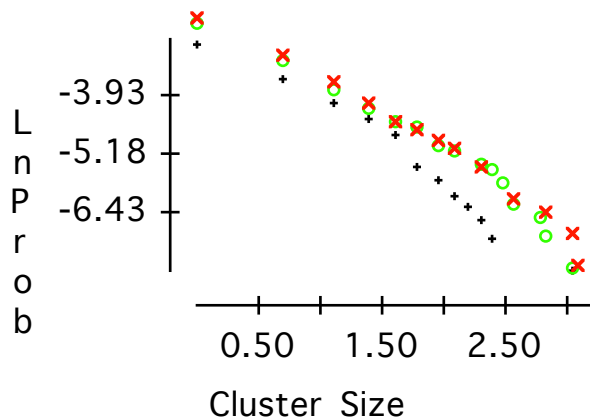

**Figure 2. Probability distribution of size of individual clusters (logarithmic) for the 3 social networks (black cross = Bluetooth, green circle = Facebook, red X = fused network).**

### Correlation between structural features

We calculated the correlation between a number of structural features of the Bluetooth and Facebook social networks. Specifically, we calculated the correlation for

- degree (0.696),
- closeness (0.555),
- betweenness (0.382), and
- clustering coefficient (0.124).

These correlations were calculated by considering all 2602 members of our cohort, and measuring their respective structural features in both the Bluetooth and Facebook graphs.

For instance, to calculate the degree correlation we measured each user's links in both datasets. In Figure 3 we see each user represented as a dot, and the vertical and horizontal axes showing the number of links that each user had in the Bluetooth and Facebook datasets. Note that significant overlap between the dots exists in the graph, especially for low-degree nodes. A similar process was followed to calculate the correlation for closeness, betweenness and clustering coefficient.

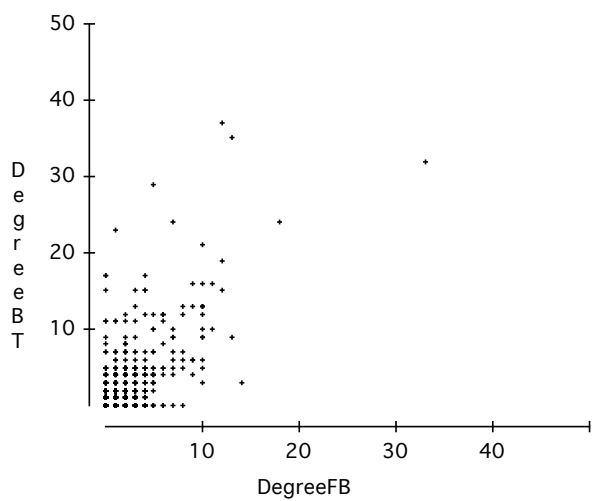

**Figure 3. Correlation of number of social ties (degree) that each user in the Bluetooth and Facebook datasets.**

## Importance of types of links

The final result we obtained from our analysis was the relative importance of the three types of links that exist in the fused social network. Recall that each social tie in the fused network may be the result of a Bluetooth encounter, Facebook friendship, or both. Figure 4 illustrates these differences by showing an excerpt of the fused social network where the ties are coloured according to their type.

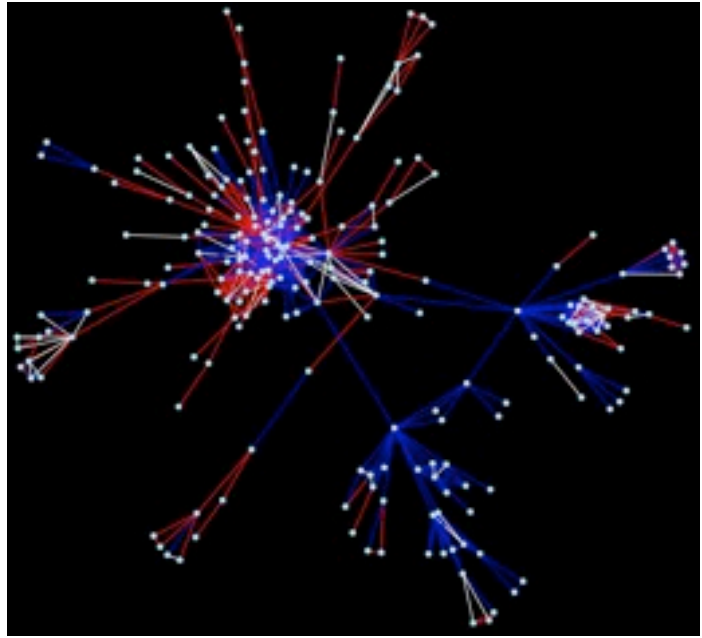

**Figure 4. The core of the fused network, with links are coloured according to their type. Blue: links resulting from Bluetooth encounters; Red: links resulting from Facebook friendship; White: links results from both Bluetooth encounters and Facebook friendship.**

Using ANOVA we identified a significant effect of link type on link betweenness (F (2,1436)= 11.212, p<0.0001). Table 1 we shows the average betweenness for each type of link in the fused network.

| Link type | Count | Mean | Std |
|---|---|---|---|
| Bluetooth (blue) | 593 | 2.46 | 2.06 |
| Facebook (red) | 440 | 2.12 | 1.99 |
| Fused (white) | 404 | 1.85 | 1.92 |

**Table 2. Summary statistics for link betweenness (logarithmic) broken down by link type. Count: the number of links for each type of link in the fused network. Mean: the average betweenness. Std: standard deviation.**

## Discussion

The Bluetooth and Facebook social networks exhibit very similar structural characteristics, suggesting that as proxies to users' actual social network they reflect similar aspects. Specifically, both networks were rather sparse with similar diameter, and both had low average degree and approximately similar average path length and clustering coefficient (see Table 1). These structural features suggest the networks are structurally similar both in terms of local characteristics (degree, clustering coefficient) as well as global characteristics (density, diameter, average path length).

It is interesting to note that while the two networks have similar characteristics, their fusion produces some interesting results. As expected, the fused network's density as well as core size significantly increase in relation to both the Bluetooth and Facebook networks. This is due to the fact that many new links are added between the same set of nodes, also evident by the increase in average degree. However, the diameter of the fused network's core does not shorten, and in fact the average path length increases. To interpret this result, we need to understand that the fused network is composed of links of three distinct types (Bluetooth, Facebook, Bluetooth+Facebook), and these types represent the effect of fusing spatial and transpatial networks. Here we observe that as users augment their traditional face-to-face spatial networks by using Facebook, they increase their local connectivity to their resulting fused network (higher average degree), but at the same time they are globally further away from everyone else in their new fused network (higher average path length) since that network is much now much larger.

We can observe just how large users' networks are by considering the cluster size we calculated in Figure 2. We note that the spatial network (Bluetooth) is made up very small clusters even though its core is larger than the Facebook network. Overall, however, the fused network's cluster size distribution is higher and almost identical to the Facebook networks. We interpret this result as suggesting that even though spatial networks can potentially offer a bigger social circle (core size), transpatial networks offer the opportunity to be part of a larger social network overall (cluster size).

Next we examine our networks from the perspective of each individual member of the cohort. Specifically, the correlation of structural measures between the Bluetooth and Facebook networks gives us insight into the relative similarity in which individuals experience their respective social circles, the opportunities in those networks, and their own role in those networks. In terms of local metrics we observe a relatively high correlation of degree (0.696), suggesting that user's local connection to their social network is similar in both spatial and transpatial networks, and that in general they make the same amount of relative effort to establish and maintain new links in each network. On the other hand, we observe a rather poor correlation of clustering coefficient (0.124), also known as transitivity. This suggests that the means by which users acquire new social ties (face-to-face vs. online) are rather independent in the extent to which the resulting ties are with people who are already friends themselves (ratio of open to closed triangles in the graph).

The correlation of global metrics follows a similar pattern, with closeness (0.555) being higher that betweenness (0.382). These results indicate that users' relative importance in their spatial and transpatial networks varies considerably (hence the low betweenness correlation), but their relative distance to everyone in the network is more similar. This latter interpretation is further supported by the similar average path length of both the Bluetooth and Facebook networks.

Overall, our correlation results indicate that some aspects of both local and global metrics appear to be similar from the perspective of individuals, while on the other hand certain local and global metrics appear to vary considerably. It is interesting to contrast this result with the global homogeneity of the Facebook and Bluetooth networks suggesting that overall the networks exhibit similar structural properties. This similarity, however, is not necessarily reflected in the way individuals utilise and experience the different ways or modalities of accessing their social networks.

Finally, an interesting result arises when we consider the impact of the various means by which social links are established on the relative importance of those links. Here we measure a link's importance by calculating its betweenness. We found a statistically significant effect of link type on link betweenness, indicating that links of spatial networks are more important than links of transpatial networks. More interestingly, we found that links that exist in both spatial and transpatial networks are of least importance. We believe these results reflect the strength of weak ties hypothesis [Granovetter, 1973]. Specifically, ties that exist in both spatial and transpatial networks are most likely with close relatives or colleagues with whom we interact closely. Such ties are not globally important in the sense that they can easily be replaced by another link in the local clique. In other words, most people that a close relative or colleague knows are people that we already know. On the other hand, we observe that relationships established solely in spatial networks are likely to be of higher importance that those relationships established in transpatial networks.

To a certain extent this result may seem counter intuitive. After all, spatial networks are ultimately bound by our ability to move in space, while transpatial networks have the potential to connect us to the whole of the world in one step. Hence, how can spatial ties have higher importance (betweenness) that transpatial ties? One explanation is that spatial networks are better at mediating the establishment of new social ties. For instance, a face-to-face meeting between A and B is also an opportunity for A to meet B's friends, since physical proximity affords this kind of natural social behavior. On the other hand, online technology has possibly not matured enough to be able to provide such affordances.

An orthogonal explanation for the importance of spatial networks is that physical co-presence, as captured by Bluetooth technology, has the potential to record "familiar strangers" relationships [Paulos & Goodman, 2004]. These are the types of relationships that users possibly do not explicitly indicate as social ties, but they can potentially activate if needed. It is not clear from literature whether the strength of social ties can be classified as strong/weak/non-existent or they follow a linear scale, but if familiar stranger ties should be present they would be classified at the low end of the linear scale. It is this presence of weak ties that our methodology reflects. A further point we should also highlight is that physical co-presence facilitates the building of trust between parties while online interactions are not necessarily as effective due to the limited channels of information they support (e.g. lack of body language and subtle communication signals).

Further evidence for the importance of spatial networks may be found in the increasing rate of urbanization, despite the availability of communication technologies that can considerably decrease cost. Major economic hubs like New York, London and Tokyo have become increasingly urbanized, even though technology now enables teleworking and remote collaboration. This increased commercial urbanization can be interpreted as a reaction to the increasing amount of available information flowing through organizations.

While our results are important in understanding the structure of spatial and transpatial networks, we can also derive a number of implications for pervasive and ubiquitous systems. Crucially, this study highlights the importance and potential of such systems as research tools, specifically in understanding community structure and collecting large amounts of data.

Furthermore, our analysis of spatial and transpatial networks can help us elicit requirements on what kind of support our pervasive systems must provide users for maintaining their social networks. We found that face-to-face ties are relatively more important in the context of the whole social network, but at the same time such networks exhibit a larger portion of small clusters. On the other hand, online mediated social ties are likely to make us part of probably larger clusters, but on average set us further away from everyone else in our network.

Further insights can be derived about developing ubiquitous systems for community maintenance and management, as opposed to systems for end users. Our results suggest that in terms of community maintenance and management, spatial and transpatial networks have similar features and properties in relation to their structure. On the other hand, ubiquitous technology for end users should allow for the differences in the roles that people have in spatial vs. transpatial networks. Our results can also provide input to the development of delay tolerant technologies, and systems that aim to capitalise on social structure for forwarding data and information. Specifically, we have identified important structural similarities and differences in how users relate to their urban and online social networks, and these can drive the dissemination and forwarding of information in a delay-tolerant fashion.

## Methodological validity

It is imperative to discuss the methodological validity of relying on Bluetooth and Facebook as social network proxies. In this study we were interested in capturing two distinct types of social networks: spatial and transpatial. Ideally, a single methodology would enable us to capture data on both networks using a single proxy, and enable us to compare our data accordingly. This was not possible in our case because a single proxy does not actually exist. Hence we utilised two distinct proxies to collect data on spatial and transpatial networks. The important issue we need to address is whether the differences we have observed in our results are due to the endemic differences of the networks we are studying, or to the differences in how the proxies we used reflect the underlying networks.

The similarity of the broad structural features of the two networks we captured, as well as the similar degree and cluster size distributions suggest that the two proxies reflected processes of similar underlying nature. This is further supported by the fact that both global and local measures display similar properties in both cases. Since as broad communities the two networks display similar characteristics, we feel confident that our correlation analyses do indeed reflect differences and similarities in our users' perspective of their social networks.

It is also interesting to point out that our two datasets consisted of slightly different types social ties: while our Facebook dataset recorded explicit friendship ties, our Bluetooth dataset recorded co-presence ties. It can be argued that this difference affected our results. A counter-argument can be made that in fact spatial networks are different from transpatial networks, hence we should expect to have different type of data. Specifically, co-presence is the crucial differentiating factor that sets apart spatial and transpatial networks, hence it is important that this is reflected in the data. While our Bluetooth data may indicate a relationship between two complete strangers who happened to be at the same place at the same time, their co-location is a significant event suggesting the possibility for social networking. As we argued earlier, humans are quite adept at forming social ties with co-located individuals, and have developed a number of conversational and linguistic mechanisms such as common ground [Clarke, 1992] to facilitate this process. Furthermore, it is not clear if there exist some tie strength threshold below which no tie should be drawn between two individuals. As part of our ongoing work we are considering more qualitative metrics for Bluetooth ties, (e.g. how often they are instantiated and for how long) as a way to offer further insight into using co-presence as a social network proxy.

## Conclusion and ongoing work

Pervasive and ubiquitous technology has the potential to act as a bridge between spatial and transpatial networks, and it is important to develop the fundamental understanding and theoretical foundations in relation to such networks. This study has provided a number of insights into the properties and relationship of spatial and transpatial social networks. Specifically, we highlight the high-level structural similarities between the two types of networks, and note the underlying differences in how individuals take part in these social networks. Furthermore, our analysis highlights the importance of spatial networks within the grand scheme of social networks.

As part of our ongoing work we are considering more qualitative metrics for Bluetooth ties, (e.g. how often they are instantiated and for how long) as a way to offer further insight into using co-presence as a social network proxy. We are also interested in exploring appropriate static representations that fully capture the temporal behaviour of the dataset.

A further aspect of our ongoing work is the development of metrics to annotate the strength of social ties. Specifically, we feel that various temporal metrics can be developed to automatically assess the strength of social ties. Furthermore, such metrics can possibly be used to derive models of tie strength, and we intend to apply these models to assess the strength of online social ties.

## Acknowledgements

The author wishes to thank Eamonn O'Neill and all other collaborators on the Cityware project.

## References

Balazinska M, Castro P (2003). Characterizing Mobility and Network Usage in a Corporate Wireless Local-Area Network. MobiSys '03: Proc. 1st Int'l Conf. on Mobile Systems, Applications and Services, ACM Press, New York, pp. 303-316.

Barabási, A.L., Albert, R. (1999). Emergence of scaling in random networks. Science, 286, 509-512.

Biggs, N., Lloyd, E. and Wilson, R. (1986). Graph Theory 1736-1936. Oxford University Press.

Booher, J. M., Chennupati, B., Onesti, N. S., and Royer, D. P. (2007). Facebook ride connect. CHI 2007 Extended Abstracts, ACM Press, New York, NY, 2043-2048.

Bott E. (1957). Family and Social Network. Roles, Norms and External Relationships in Ordinary Urban Families. London, Tavistock Publishers.

Castells, M. (1995). The Information City. Oxford, Blackwell.

Chaintreau A, Hui P, Crowcroft J, Diot C, Gass R, Scott J (2006). Impact of Human Mobility on the Design of Opportunistic Forwarding Algorithms. Proc. 25th IEEE Conf. on Computer Communications (INFOCOM), IEEE CS Press, New York, NY, USA, 2006.

Clark, H.H. (1992). Arenas of language use, Chicago, University of Chicago Press.

Eagle N, Pentland A (2006). Reality mining: sensing complex social systems. Personal and Ubiquitous Computing, 10(4):255-268, Springer-Verlag, London.

Foth, M. (2006). Facilitating Social Networking in Inner-City Neighborhoods. Computer 39, 9 (Sep. 2006), 44-50.

Freeman, L. (2004). The Development of Social Network Analysis: A Study in the Sociology of Science. Vancouver, BC, Canada: Empirical Press.

Gluckman, M. & Aronoff, M. J. (1976). Freedom and constraint : a memorial tribute to Max Gluckman. Assen: Van Gorcum.

Granovetter, M. (1973). The strength of weak ties. American Journal of Sociology, 78(6), 1360-1380.

Herrmann, T., Kunau, G., Loser, K., and Menold, N. (2004). Socio-technical walkthrough: designing technology along work processes. Proc. Conference on Participatory Design (PDC), ACM, New York, NY, 132-141.

Kostakos, V. and O'Neill E. (2008). Cityware: Urban Computing to Bridge Online and Real-world Social Networks. In M. Foth (Ed.), Handbook of Research on Urban Informatics: The Practice and Promise of the Real-Time City. Hershey, PA: Information Science Reference, IGI Global, pp. 195-204.

Kostakos, V., O'Neill, E., and Penn, A. (2006). Designing Urban Pervasive Systems. Computer 39, 9 (Sep. 2006), 52-59.

Lampe, C. A., Ellison, N., and Steinfield, C. (2007). A familiar face(book): profile elements as signals in an online social network. Proc. SIGCHI Conference on Human Factors in Computing Systems (CHI), ACM Press, New York, NY, 435-444.

Lee, K. M. and Nass, C. 2003. Designing social presence of social actors in human computer interaction. Proc. SIGCHI Conference on Human Factors in Computing Systems (CHI), ACM, New York, NY, 289-296.

McNett M, Voelker GM (2005) Access and Mobility of Wireless PDA Users. SIGMOBILE Mob. Comput. Commun. Rev., 9(2):40-55.

Millen, D. R. and Patterson, J. F. (2002). Stimulating social engagement in a community network. Proc. Conference on Computer Supported Cooperative Work (CSCW), ACM, New York, NY, 306-313.

Nicolai T, Yoneki E, Behrens N, Kenn H (2005) Exploring Social Context with the Wireless Rope. On the Move to Meaningful Internet Systems 2006: OTM 2006 Workshops, Part I, LNCS 4277:874-883, Springer Berlin/Heidelberg.

Paulos, E. and Goodman, E. (2004). The familiar stranger: anxiety, comfort, and play in public places. Proceedings of the SIGCHI conference on Human factors in computing systems, p.223-230, April 24-29, 2004, Vienna, Austria.

Riegelsberger, J. and Vasalou, A. (2007). Trust 2.1: advancing the trust debate. Proc. SIGCHI Conference on Human Factors in Computing Systems (CHI),Extended Abstracts, ACM Press, New York, NY, 2137-2140.

Watts, D.J., Strogatz, S.H. (1998). Collective dynamics of small-world networks. Nature, 393, 440.

Wellman, B. (1979). The Community Question: The Intimate Networks of East Yorkers. American Journal of Sociology 84, March, 1979: 1201-31.

United Nations (2007). World Urbanization Prospects: The 205 Revision. See http://esa.un.org/unup/, last access August 28, 2007.